\documentclass[aps,prd,reprint,superscriptaddress,nofootinbib,floatfix]{revtex4-2}

\usepackage{amsmath,amssymb,bm}
\usepackage{graphicx}
\usepackage{booktabs}
\usepackage{multirow}
\usepackage{xcolor}
\usepackage{microtype}
\usepackage{siunitx}
\usepackage{hyperref}
\usepackage{url}

\hypersetup{
  colorlinks=true,
  linkcolor=blue!55!black,
  citecolor=blue!55!black,
  urlcolor=blue!55!black
}
\graphicspath{{figures/}}

\newcommand{\gD}{\gamma_D}
\newcommand{\met}{E_T^{\rm miss}}
\newcommand{\ptg}{p_T^\gamma}
\newcommand{\BR}{\mathcal{B}}
\newcommand{\gev}{\mathrm{GeV}}
\newcommand{\tev}{\mathrm{TeV}}
\newcommand{\fb}{\mathrm{fb}}
\newcommand{\CLs}{\mathrm{CL}_s}
\newcommand{\rev}[1]{#1}
\newcommand{\audit}[1]{#1}

\begin{document}

\title{\audit{Probing Massive Invisible Dark Photons in Higgs Decays via
Gluon--Gluon Fusion at ATLAS Run 3}}

\author{Waqas Ahmed}
\email{waqasmit@hbpu.edu.cn}
\affiliation{Center for Fundamental Physics and School of Artificial Intelligence,
Hubei Polytechnic University, Huangshi 435003, China}

\author{Ammara Ahmad}
\email{ammara.ahmad@ku.ac.ae}
\affiliation{Khalifa University of Science and Technology, Abu Dhabi, United Arab Emirates}
\author{Shabbar Raza}
\email{shabbar.raza@fuuast.edu.pk}
\affiliation{Department of Physics, Federal Urdu University of Arts, Science and Technology, Karachi 75300, Pakistan}


\begin{abstract}
The ATLAS Collaboration has searched for Higgs-boson decays into a photon and missing transverse momentum using $135~\mathrm{fb}^{-1}$ of Run~3 proton--proton collision data at $\sqrt{s}=13.6~\mathrm{TeV}$. For the benchmark decay $H\to\gamma\gD$ with a massless invisible dark photon, ATLAS reports observed (expected) 95\% confidence-level limits of $1.4\%$ ($1.2\%$) on the branching fraction. We reinterpret this search for a massive detector-invisible dark photon with $0\leq m_{\gD}\leq70~\gev$, focusing on Higgs production through gluon--gluon fusion. \audit{The relative signal acceptance, normalized to the massless benchmark, remains nearly unchanged up to $m_{\gD}\simeq40~\gev$, but decreases to $0.888$, $0.736$, and $0.524$ at $50$, $60$, and $70~\gev$, respectively, as the photon-$p_T$ and transverse-mass spectra move toward the analysis thresholds.} Using the published massless gluon--gluon-fusion yield for normalization, we obtain simplified expected (observed) 95\% confidence-level limits on $\BR(H\to\gamma\gD)$ from $1.64\%$ ($1.76\%$) at $m_{\gD}=30~\gev$ to $3.50\%$ ($3.75\%$) at $70~\gev$. These results extend the ATLAS interpretation to massive invisible dark photons and quantify the resulting loss of sensitivity at large dark-photon mass.
\end{abstract}

\maketitle

\section{Introduction}
\label{sec:intro}

The Higgs boson is a sensitive probe of physics beyond the Standard Model (SM). 
Its interactions can provide a portal to new particles that are neutral under the SM gauge group and only weakly coupled to visible matter. 
Current measurements of Higgs production and decay are in good agreement with SM predictions. 
At the same time, they still allow room for rare or exotic Higgs decays, with the allowed branching fractions depending on the final state and on the assumptions entering global fits 
\cite{ATLAS_HiggsMap2022,CMS_HiggsPortrait2022,Curtin2014,Cepeda2022}. 
Such decays therefore provide a well-motivated way to search for dark sectors, feebly interacting particles, and other new states that may evade conventional resonance searches 
\cite{Davoudiasl2013,Curtin2015DarkPhotons,ArkaniHamed2009,Zurek2014,Beacham2020,FIPs2020,FIPs2022}.

\rev{The vector-portal idea itself predates the modern dark-photon terminology. Additional Abelian gauge factors and non-orthogonal $U(1)$ sectors were considered in early work, and kinetic mixing subsequently provided the standard renormalizable portal between visible and hidden gauge sectors \cite{Okun1982,GalisonManohar1984,Holdom1986}. The phenomenology of light hidden vectors was later developed systematically for low-energy colliders, fixed-target experiments, and dark-sector searches \cite{EssigSchusterToro2009,Bjorken2009,Batell2009,Ilten2018,Fabbrichesi2020}. This history is useful for the present analysis because a massive invisible vector need not be identified with one unique ultraviolet realization of kinetic mixing.}

\audit{A particularly economical hidden-sector extension introduces an additional Abelian gauge symmetry, $U(1)_D$. Kinetic mixing between two Abelian field strengths is a renormalizable interaction connecting the visible and dark sectors \cite{Holdom1986,Dienes1997,Fabbrichesi2020}. For a massive dark photon, the mixing generally induces physical couplings to the Standard Model electromagnetic current. For an exactly massless dark photon, the observable consequences depend on the matter charge assignments and on the choice of gauge-field basis. Higgs decays involving a photon and a dark photon can instead be generated by loop-induced or higher-dimensional interactions associated with states that connect the two sectors \cite{Gabrielli2014,Biswas2016,BeauchesneChiang2023PRD}. The resulting dark-photon phenomenology depends on the vector mass, its couplings, and the spectrum of additional dark-sector states. The vector may decay promptly to Standard Model particles, produce displaced visible decays, decay invisibly into lighter dark-sector states, or be sufficiently long lived to escape the detector. These possibilities motivate complementary visible- and invisible-decay searches across fixed-target, flavor-factory, collider, and intensity-frontier experiments \cite{Essig2013,Alexander2016,Battaglieri2017,Beacham2020,FIPs2020,FIPs2022,Harris2022,BaBarInvisible2017,NA64Invisible2018,NA64MissingEnergy2019,BaBarVisible2014,LHCbDarkPhoton2020,NA64Latest2023}.}

The decay
\begin{equation}
H\to\gamma\gD
\label{eq:signal_decay}
\end{equation}
occupies a distinctive position within this program. It yields a semi-visible final state in which a hard photon tags an invisible degree of freedom, while the Higgs resonance fixes the underlying two-body scale. The process can be generated by loop-induced operators involving messenger fields charged under both the SM and dark gauge groups, and it has been investigated in effective descriptions and ultraviolet completions in Refs.~\cite{Gabrielli2014,Biswas2016,Biswas2017,Biswas2022Review,BeauchesneChiang2023PRL,BeauchesneChiang2023PRD,KroviLowZhang2020}. More generally, photon-plus-invisible or semi-visible Higgs signatures can also arise from low-scale supersymmetry breaking and other semi-dark Higgs constructions \cite{Petersson2012,AguilarSaavedra2022}. The same reconstructed topology is therefore useful beyond any single dark-photon ultraviolet completion.

\rev{Closely related studies have also considered associated Higgs--dark-photon production at lepton colliders, the extraction of a massive dark photon's properties from semi-invisible Higgs decays, and Higgs observables generated by inert electroweak multiplets or other non-decoupling charged states \cite{Biswas2015,BeauchesneChiang2022,BeauchesneChiang2023Inert,Banta2022}. These works reinforce the point that the photon-plus-invisible Higgs channel probes both the existence of a dark sector and, when sufficient kinematic information is available, properties of the invisible state.}

For a massless invisible vector, Eq.~\eqref{eq:signal_decay} produces a monochromatic photon of energy $m_H/2$ in the Higgs rest frame. The collider phenomenology of this benchmark has been developed in Refs.~\cite{Gabrielli2014,Biswas2016,Biswas2017,Biswas2022Review}, while theoretical studies of mediator sectors and consistency constraints have clarified the circumstances under which branching fractions at experimentally interesting levels can arise \cite{BeauchesneChiang2023PRL,BeauchesneChiang2023PRD}. Searches at hadron colliders have targeted the same semi-visible decay in several Higgs-production modes. Earlier analyses considered associated $ZH$ production and vector-boson fusion (VBF) \cite{CMSZH2019,CMSVBF2021,ATLASVBF2022,ATLASZH2023}, and the corresponding ATLAS Run~2 combination reached percent-level sensitivity for the massless benchmark \cite{ATLASCombination2024}. Photon-plus-missing-momentum searches designed for more general dark-matter or exotic signatures provide closely related experimental context \cite{CMSHiggsPhotonInvisible2016,CMSMonophoton2017,ATLASMonophoton2017,ATLASMonophoton2021}.

\rev{Importantly, a massive interpretation is not without precedent: the Run~2 ATLAS $ZH$ analysis explicitly considered dark-photon masses up to $40~\gev$ \cite{ATLASZH2023}, and dedicated phenomenological work has studied how a nonzero invisible mass could be inferred from Higgs-decay kinematics \cite{BeauchesneChiang2022}. The specific unresolved question addressed here is therefore narrower: how the new low-threshold Run~3 gluon--gluon fusion (ggF) selection responds to a massive invisible vector, and how far that response can be extrapolated using only public information.}

\audit{Run~3 provides a particularly relevant setting for Higgs production through ggF. The ATLAS analysis uses a dedicated trigger that combines the photon transverse momentum, missing transverse momentum, and the transverse mass of the photon--missing-momentum system, enabling a low-threshold selection for this topology. Using $135~\fb^{-1}$ of proton--proton collision data at $\sqrt{s}=13.6~\tev$, the search finds no significant excess and reports an observed (expected) 95\% confidence-level upper limit of $1.4\%$ ($1.2\%$) on $\BR(H\to\gamma\gD)$ for a massless dark photon. Combining the Run~3 result with previous Run~2 searches gives an observed (expected) limit of approximately $0.9\%$ ($0.9\%$) \cite{ATLASRun3DarkPhoton2026}. Since ggF provides the largest selected signal component for the massless benchmark, it is a natural starting point for a public mass-dependent reinterpretation.}

The published benchmark nevertheless addresses only the special point $m_{\gD}=0$. For a massive invisible daughter, the two-body decay remains open throughout
\begin{equation}
0\leq m_{\gD}<m_H,
\end{equation}
but the event kinematics change continuously with $m_{\gD}$. In particular, the photon becomes softer in the Higgs rest frame, the missing-particle energy changes, and the conventional transverse-mass variable employed by the experimental analysis is increasingly compressed. These changes are especially important because the Run~3 search deliberately operates near relatively low fixed thresholds. Consequently, a massive interpretation cannot be obtained reliably by multiplying the massless limit by a decay phase-space factor: the detector-level acceptance must be reevaluated.

This observation is also conceptually distinct from the usual discussion of massive kinetically mixed dark photons. In the present work, $\gD$ denotes a spin-one state that is invisible on detector scales. This can correspond to a sufficiently long-lived vector with extremely suppressed visible couplings, or to a vector whose decays are dominantly into lighter invisible dark-sector states. We do not attempt to map the result onto a unique kinetic-mixing parameter because that translation is model dependent and can involve the dark gauge coupling, additional particle masses, and the dark-photon lifetime \cite{Holdom1986,Pospelov2009,Fabbrichesi2020,BentoHaberSilva2024,BaBarInvisible2017,NA64Invisible2018}. Instead, the experimentally robust quantity is the mass-dependent limit on the Higgs branching fraction into a photon plus a detector-invisible vector.

\rev{The present paper addresses this specific public-recast gap for ggF production.} We simulate eight masses,
\begin{equation}
 m_{\gD}=0,10,20,30,40,50,60,70~\gev,
\end{equation}
using a common hard-event sample, matched matrix elements, parton showering, and fast detector simulation. The recastable part of the Run~3 signal-region selection is then applied identically to all masses. Rather than relying on an absolute fast-simulation efficiency, we use the acceptance ratio
\begin{equation}
R_A(m_{\gD})=\frac{A(m_{\gD})}{A(0)}
\end{equation}
and anchor the normalization to the public massless ggF signal yield. \audit{This ratio strategy follows the general logic of collider reinterpretation: it isolates the parameter dependence that can be robustly recalculated while inheriting experimental information that cannot be reconstructed faithfully from public detector-level inputs alone \cite{CranmerYavin2011,LHCReinterpretation2020}.}

\audit{Our main result is that the relative ggF acceptance, $R_A(m_{\gD})$, is approximately mass independent through $m_{\gD}\simeq40~\gev$ within the available Monte Carlo precision, but decreases rapidly once the rest-frame photon momentum approaches the fixed photon threshold.} The effect becomes pronounced at $50$--$70~\gev$, where the selected events increasingly depend on recoil of the Higgs boson against QCD radiation. We propagate the mass dependence into a simplified one-bin profile-likelihood model normalized to the published Run~3 yields and provide expected and observed branching-fraction limits across the scan. \audit{We also perform a defined selection-variation stress test to quantify how strongly the acceptance ratio depends on the recastable analysis thresholds.}

The scope is intentionally narrower than an experimental reanalysis. We do not simulate VBF, $WH$, or $ZH$ production, do not reconstruct ATLAS-only observables without a faithful fast-simulation analogue, and do not attempt to reproduce the full multibin signal/control-region likelihood. The resulting limits should therefore be read as a transparent ggF-only public recast. This distinction is particularly important at high $m_{\gD}$, where the signal is selected increasingly from the boosted ggF recoil tail and the leading-order matched simulation should ultimately be checked against higher-order descriptions such as NNLOPS \cite{Hamilton2013,deFlorian2017,Karlberg2024}.

The paper is organized as follows. Section~\ref{sec:framework} defines the phenomenological signal model and the mass-dependent two-body kinematics. Section~\ref{sec:atlasinputs} summarizes the public Run~3 inputs and the precise scope of the recast. Section~\ref{sec:simulation} describes event generation, matching, showering, and fast detector simulation. Validation and robustness tests are given in Sec.~\ref{sec:validation}. Section~\ref{sec:acceptance} presents the mass-dependent acceptance and transverse-mass migration. The simplified statistical interpretation and branching-ratio limits are developed in Sec.~\ref{sec:stats}. The theoretical and experimental limitations, together with the path toward an all-production-mode reinterpretation, are discussed in Sec.~\ref{sec:discussion}, followed by our conclusions in Sec.~\ref{sec:conclusions}.

\section{Phenomenological framework and massive-vector kinematics}
\label{sec:framework}

We consider a neutral spin-one state $\gD$ produced in the exotic Higgs decay of Eq.~\eqref{eq:signal_decay}. The state is assumed to be invisible to the ATLAS detector on the event scale relevant to the search. At the phenomenological level this means that its four-momentum contributes to $\met$ but that no reconstructed visible decay products are assigned to $\gD$.

\audit{For a massive dark photon, the experimental signature depends on its couplings and decay channels. If it couples appreciably to Standard Model fermions through kinetic mixing, visible decays can occur \cite{Holdom1986,Fabbrichesi2020}. A detector-invisible signature can instead arise from sufficiently weak visible couplings, a decay length larger than the detector dimensions, or dominant decays into lighter dark-sector states \cite{BaBarInvisible2017,NA64Invisible2018,NA64MissingEnergy2019}. Since the relation among the dark-photon mass, lifetime, visible couplings, and invisible branching fractions depends on the underlying model, we do not impose a specific microscopic realization. We therefore treat $\gD$ phenomenologically as a detector-invisible state and present the results directly in the $(m_{\gD},\BR)$ plane.}

A representative effective interaction after electroweak symmetry breaking 
can be written schematically as
\begin{equation}
\mathcal{L}_{\rm eff}
\supset
\frac{c_{\gamma D}}{v}\,
H F_{\mu\nu}X^{\mu\nu},
\label{eq:effop}
\end{equation}
where $F_{\mu\nu}$ and $X_{\mu\nu}$ denote the electromagnetic and
dark-vector field strengths, respectively, $v\simeq246~\gev$ is the
electroweak vacuum expectation value, and $c_{\gamma D}$ parametrizes
the short-distance dynamics generating the effective interaction
\cite{Gabrielli2014,Biswas2016,BeauchesneChiang2022}.
Gauge-invariant realizations above the electroweak scale can be written
in terms of operators involving the Higgs doublet and the electroweak
field strengths, which reduce to Eq.~\eqref{eq:effop} after electroweak
symmetry breaking \cite{BeauchesneChiang2022}.
The precise normalization of the effective coupling is not required
below, since the collider interpretation is expressed directly in
terms of $\BR(H\to\gamma\gD)$.

For an interaction of the form in Eq.~\eqref{eq:effop}, the two-body
partial width contains the characteristic kinematic factor
\begin{equation}
\Gamma(H\to\gamma\gD)
\propto
\left(1-\frac{m_{\gD}^{2}}{m_H^{2}}\right)^{3},
\label{eq:phasewidth}
\end{equation}
up to the normalization of the effective coupling
\cite{BeauchesneChiang2022}.
The factor in Eq.~\eqref{eq:phasewidth} describes the suppression of
the decay width at fixed coupling. In contrast, the present analysis
focuses on the separate mass dependence of the experimental acceptance
for a fixed value of $\BR(H\to\gamma\gD)$.

\subsection{Two-body decay scales}

The kinematics of a two-body decay are completely fixed in the rest frame
of the parent particle by the daughter masses
\cite{PDG2024,BeauchesneChiang2022}.
For an on-shell Higgs boson decaying into a massless photon and a dark
vector of mass $m_{\gD}$, the two daughter momenta have a common magnitude
\begin{equation}
 p_\gamma^*=p_{\gD}^*
 =\frac{m_H^2-m_{\gD}^2}{2m_H}.
\label{eq:pstar}
\end{equation}
The corresponding dark-vector energy is
\begin{equation}
 E_{\gD}^*
 =\frac{m_H^2+m_{\gD}^2}{2m_H}.
\label{eq:ED}
\end{equation}
Thus, increasing $m_{\gD}$ reduces the photon energy in the Higgs rest
frame while increasing the energy carried by the invisible state.

The ATLAS searches for $H\to\gamma\gD$ characterize the
photon--missing-momentum system using the conventional transverse mass
\cite{ATLASZH2023,ATLASRun3DarkPhoton2026},
\begin{equation}
 m_T
 =\sqrt{
 2\ptg\met
 \left[
 1-\cos\Delta\phi
 \bigl(\gamma,\bm{p}_T^{\rm miss}\bigr)
 \right]
 }.
\label{eq:mtdef}
\end{equation}
For a massless invisible particle, this variable provides the usual
transverse-mass description of the recoil system. For a massive $\gD$,
the experimental definition remains unchanged, although it no longer
coincides with the transverse mass constructed using the true nonzero
invisible-particle mass.

In the idealized limit in which the Higgs boson has negligible transverse
recoil, the upper kinematic scale of the massless-proxy transverse mass is
\begin{equation}
 m_T^{\rm edge}
 \simeq 2p_\gamma^*
 =\frac{m_H^2-m_{\gD}^2}{m_H}.
\label{eq:mtedge}
\end{equation}
This relation follows directly from the two-body kinematics above and
shows that both the characteristic photon momentum and the transverse-mass
endpoint decrease as $m_{\gD}$ increases. Consequently, the signal
distributions move progressively toward the fixed photon-$p_T$ and
$m_T$ thresholds used in the ATLAS selection
\cite{ATLASRun3DarkPhoton2026}.

\begin{table}[t]
\caption{Two-body kinematic scales for the simulated dark-photon masses, evaluated with $m_H=125~\gev$.}
\label{tab:kinematics}
\centering
\begin{ruledtabular}
\begin{tabular}{c c c}
$m_{\gD}$ [GeV] & $p_\gamma^*$ [GeV] & $m_T^{\rm edge}$ [GeV]\\
\hline
0  & 62.5 & 125.0\\
10 & 62.1 & 124.2\\
20 & 60.9 & 121.8\\
30 & 58.9 & 117.8\\
40 & 56.1 & 112.2\\
50 & 52.5 & 105.0\\
60 & 48.1 & 96.2\\
70 & 42.9 & 85.8\\
\end{tabular}
\end{ruledtabular}
\end{table}

\rev{The analytic scales in Table~\ref{tab:kinematics} are used below as a guide to the fully simulated distributions. Rather than adding a separate derived kinematic-scale plot, we display the reconstructed photon-$p_T$ and $m_T$ threshold migration directly in Sec.~\ref{sec:acceptance}.}

\section{Public ATLAS inputs and scope of the recast}
\label{sec:atlasinputs}

The Run~3 search in Ref.~\cite{ATLASRun3DarkPhoton2026} is based on $135~\fb^{-1}$ of $pp$ collisions at $\sqrt{s}=13.6~\tev$. It employs an inclusive photon-plus-missing-momentum signal region and several control regions in a simultaneous profile-likelihood fit. The signal-region requirements include an analysis-trigger photon, $\met>100~\gev$, $m_T>80~\gev$, exactly one baseline photon with $\ptg>50~\gev$, a lepton veto, no more than three central jets, and a $b$-jet veto. Additional selections use the photon pseudorapidity, the photon--missing-momentum azimuth, the vector sum of selected central jets, and the two leading jets. The experimental analysis also employs a hard-scatter-vertex discriminant, an object-based missing-momentum significance, and a no-JVT missing-momentum quantity that do not have faithful one-to-one counterparts in the fast simulation used here.

The public signal composition after the full ATLAS selection is approximately $68\%$ ggF, $23\%$ VBF, $6\%$ $WH$, and $3\%$ $ZH$. For a massless dark photon and $\BR(H\to\gamma\gD)=1\%$, the published ggF yield in the signal region is
\begin{equation}
N_{\rm ggF}^{0}=166\pm24.
\label{eq:ggFanchor}
\end{equation}
The observed signal-region count and the post-fit background prediction are
\begin{equation}
 n_{\rm obs}=11327,
 \qquad
 b=11298\pm101.
\label{eq:publiccounts}
\end{equation}
These quantities are the normalization and statistical anchors of the reinterpretation.

No background sample is generated in the present work. This is a deliberate choice. The ATLAS background model includes prompt electroweak photon processes, jets and electrons misidentified as photons, and instrumental missing momentum, with important components constrained directly by control-region data. A standalone Delphes background campaign would not reproduce those data-driven constraints and would therefore provide a less faithful normalization than the published post-fit estimate. The purpose here is instead to vary the massive signal hypothesis while inheriting the experimental background information.

Table~\ref{tab:selection} makes explicit which parts of the public signal-region definition are reproduced event by event and which are inherited through the massless anchor. This separation is essential to the interpretation of the result. We assume that the efficiency of the unavailable requirements has no strong additional dependence on $m_{\gD}$ after conditioning on the reconstructed kinematics included in the recast. The assumption is most credible for modest masses and becomes progressively less secure when the signal is accepted primarily from the tails of the Higgs-recoil distribution.

\begin{table*}[t]
\caption{Mapping of the public Run~3 signal-region requirements to the present fast-simulation recast. ``Anchored'' denotes an efficiency inherited from the published massless ggF yield rather than emulated event by event.}
\label{tab:selection}
\centering
\footnotesize
\begin{tabular}{@{}lll@{}}
\toprule
\parbox[t]{0.23\textwidth}{\textbf{ATLAS requirement}} &
\parbox[t]{0.42\textwidth}{\textbf{Recast implementation}} &
\parbox[t]{0.25\textwidth}{\textbf{Status and role}}\\
\midrule
\parbox[t]{0.23\textwidth}{Analysis photon trigger} & \parbox[t]{0.42\textwidth}{Not emulated} & \parbox[t]{0.25\textwidth}{Anchored to the massless ggF yield}\\[1mm]
\parbox[t]{0.23\textwidth}{$\met>100~\gev$} & \parbox[t]{0.42\textwidth}{Delphes missing transverse momentum} & \parbox[t]{0.25\textwidth}{Implemented}\\[1mm]
\parbox[t]{0.23\textwidth}{$m_T(\gamma,\met)>80~\gev$} & \parbox[t]{0.42\textwidth}{Eq.~\eqref{eq:mtdef}} & \parbox[t]{0.25\textwidth}{Implemented}\\[1mm]
\parbox[t]{0.23\textwidth}{Exactly one baseline photon} & \parbox[t]{0.42\textwidth}{Exactly one Delphes photon satisfying baseline object definition} & \parbox[t]{0.25\textwidth}{Implemented approximately}\\[1mm]
\parbox[t]{0.23\textwidth}{$\ptg>50~\gev$, $|\eta_\gamma|<1.75$} & \parbox[t]{0.42\textwidth}{Leading baseline photon} & \parbox[t]{0.25\textwidth}{Implemented}\\[1mm]
\parbox[t]{0.23\textwidth}{Electron, muon, and hadronic-$\tau$ veto} & \parbox[t]{0.42\textwidth}{Delphes reconstructed-object veto} & \parbox[t]{0.25\textwidth}{Implemented approximately}\\[1mm]
\parbox[t]{0.23\textwidth}{At most three central jets} & \parbox[t]{0.42\textwidth}{anti-$k_T$ jets with $p_T>20~\gev$ and $|\eta|<2.5$} & \parbox[t]{0.25\textwidth}{Implemented}\\[1mm]
\parbox[t]{0.23\textwidth}{$b$-jet veto} & \parbox[t]{0.42\textwidth}{Delphes $b$-tag decision} & \parbox[t]{0.25\textwidth}{Implemented approximately}\\[1mm]
\parbox[t]{0.23\textwidth}{Hard-scatter vertex discriminant} & \parbox[t]{0.42\textwidth}{No faithful fast-simulation analogue} & \parbox[t]{0.25\textwidth}{Anchored}\\[1mm]
\parbox[t]{0.23\textwidth}{Object-based $\met$ significance} & \parbox[t]{0.42\textwidth}{No faithful fast-simulation analogue} & \parbox[t]{0.25\textwidth}{Anchored}\\[1mm]
\parbox[t]{0.23\textwidth}{No-JVT missing-momentum requirement} & \parbox[t]{0.42\textwidth}{No faithful fast-simulation analogue} & \parbox[t]{0.25\textwidth}{Anchored}\\[1mm]
\parbox[t]{0.23\textwidth}{$\Delta\phi(\met,-\gamma)>1.25$} & \parbox[t]{0.42\textwidth}{Reconstructed photon and missing-momentum vectors} & \parbox[t]{0.25\textwidth}{Implemented}\\[1mm]
\parbox[t]{0.23\textwidth}{$\Delta\phi(\met,-\sum\bm{p}_T^{\rm jets})<0.75$} & \parbox[t]{0.42\textwidth}{Vector sum of selected central jets; undefined angle fails nominally} & \parbox[t]{0.25\textwidth}{Implemented approximately}\\[1mm]
\parbox[t]{0.23\textwidth}{$\Delta\phi(j_1,j_2)<2.5$} & \parbox[t]{0.42\textwidth}{Two leading central jets; events with fewer than two pass} & \parbox[t]{0.25\textwidth}{Implemented approximately}\\
\bottomrule
\end{tabular}
\end{table*}

\section{Event generation and detector simulation}
\label{sec:simulation}

Gluon-fusion Higgs production is generated at $\sqrt{s}=13.6~\tev$ with \textsc{MadGraph5\_aMC@NLO} 3.6.4 \cite{Alwall2014}. The hard process is treated in the heavy-top effective interaction, and zero- and one-parton leading-order matrix elements are combined before showering. The use of a matched 0+1 parton sample is important for the present problem because the high-$m_{\gD}$ acceptance depends increasingly on transverse recoil of the Higgs boson.

A common sample of $200000$ pre-matching hard events is used for all eight dark-photon masses. At each mass point the Higgs decay is imposed in \textsc{Pythia 8} \cite{Bierlich2022}, with the dark photon assigned PDG identifier 4900022 and kept stable. MLM-style matrix-element/parton-shower matching is used to avoid double counting of hard radiation between the matrix element and the shower \cite{Mangano2007}. Between $89102$ and $90007$ events survive matching depending on the mass point and shower seed.

\audit{The common hard-event sample keeps the pre-decay Higgs-production kinematics identical across the mass hypotheses before the independent shower and detector passes. This reduces spurious cross-mass differences in the generated production spectrum, but the analysis does not exploit event-level correlations statistically. Because independent shower seeds are used and the cross-mass covariance is not retained, the acceptance-ratio uncertainties below are evaluated with zero covariance.}

\subsection{Parton shower, detector response, and object reconstruction}

Showered and hadronized events are passed through \textsc{Delphes 3} \cite{Delphes2014}. Jets are reconstructed with the anti-$k_T$ algorithm \cite{Cacciari2008}. The same detector card, object definitions, analysis code, and selection are used for every mass; only $m_{\gD}$ and the independent shower seed vary.

For each matched event sample we record the cumulative cutflow, the four $m_T$ intervals employed by the public Run~3 analysis,
\begin{equation}
\begin{aligned}
80&<m_T<110, & 110&<m_T<140,\\
140&<m_T<200, & m_T&\geq200~\gev,
\end{aligned}
\label{eq:mtbins}
\end{equation}
and the selected $m_T$, $\met$, and $\ptg$ distributions.

The implemented efficiency is
\begin{equation}
A(m_{\gD})=\frac{N_{\rm sel}(m_{\gD})}{N_{\rm match}(m_{\gD})},
\label{eq:acceptance}
\end{equation}
and the central observable used for extrapolation is
\begin{equation}
R_A(m_{\gD})=\frac{A(m_{\gD})}{A(0)}.
\label{eq:ratio}
\end{equation}
The absolute Delphes efficiency is never used to normalize the ATLAS signal. Instead, Eq.~\eqref{eq:ratio} is multiplied by the public yield in Eq.~\eqref{eq:ggFanchor}. Common detector mismodeling therefore cancels to first approximation in the mass ratio.

\audit{The Monte Carlo statistical uncertainty is estimated from the standard binomial counting variance, with first-order uncertainty propagation used for the acceptance ratio \cite{PDG2024}.} For $m_{\gD}>0$,
\begin{equation}
\sigma_A^2\simeq\frac{A(1-A)}{N_{\rm match}},
\end{equation}
and the ratio uncertainty is evaluated as
\begin{equation}
\left(\frac{\sigma_{R_A}}{R_A}\right)^2
\simeq
\left(\frac{\sigma_{A(m)}}{A(m)}\right)^2
+
\left(\frac{\sigma_{A(0)}}{A(0)}\right)^2.
\label{eq:ratioerr}
\end{equation}
\audit{Because the event-level cross-mass covariance is not retained, it is set to zero in the uncertainty estimate. If the shared hard events induce a positive covariance, this prescription overestimates the uncertainty on the ratio and is therefore conservative under that expectation.}

\section{Validation and robustness strategy}
\label{sec:validation}

\begin{figure}[t]
\centering
\includegraphics[width=\columnwidth]{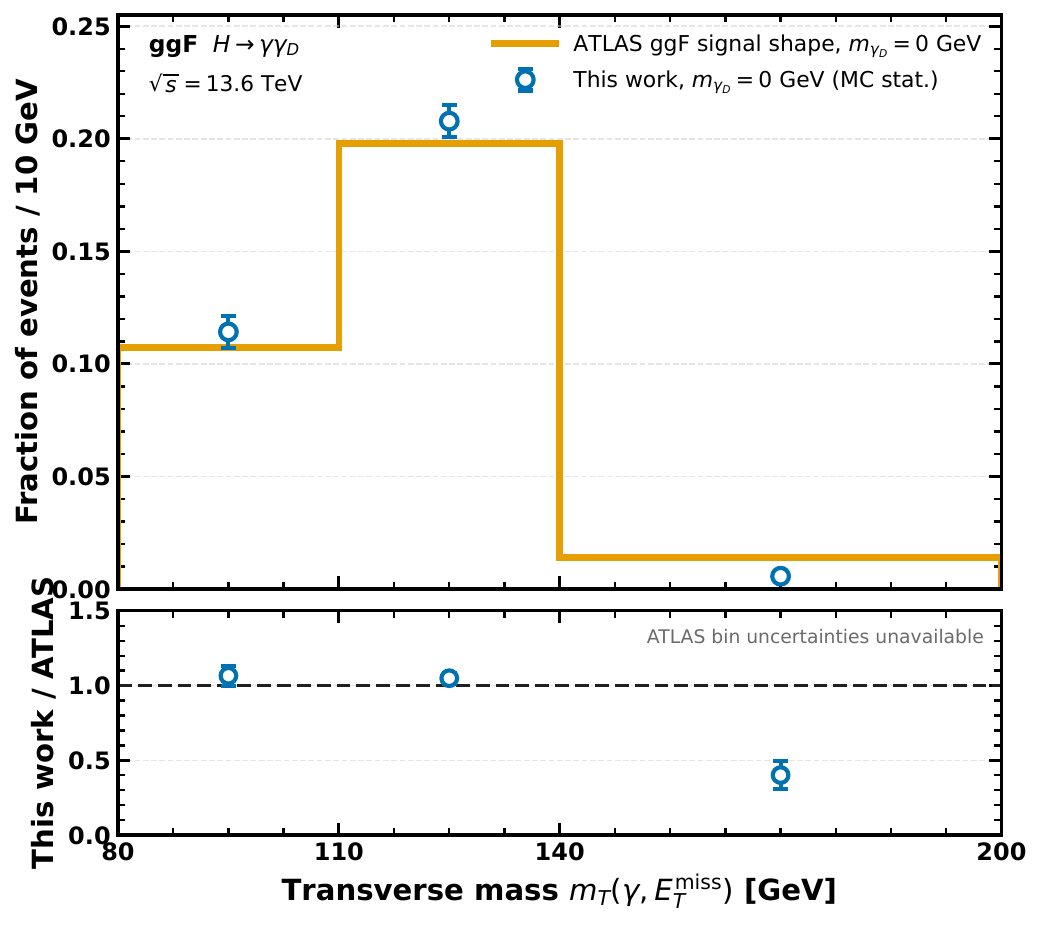}
\caption{\audit{Massless-benchmark validation of the transverse-mass shape. The Delphes ggF sample and digitized public ATLAS ggF template are independently normalized in $80\leq m_T<200~\gev$ after the cuts common to the two descriptions. The lower panel shows their ratio. Statistical uncertainties are shown for the recast sample; ATLAS bin-by-bin uncertainties are not publicly available for this comparison. The plot is therefore a shape-closure diagnostic, while the absolute normalization is inherited from Eq.~\eqref{eq:ggFanchor}.}}
\label{fig:validation}
\end{figure}

\audit{The first validation uses the massless benchmark and compares the normalized ggF $m_T$ shape after the cuts common to the two descriptions with the corresponding public ATLAS ggF template, as shown in Fig.~\ref{fig:validation}. The recast follows the main shape of the populated part of the public template. The high-$m_T$ tail contains limited Monte Carlo statistics, and the ATLAS bin-by-bin uncertainties needed for a quantitative goodness-of-fit test are not publicly available. We therefore use this comparison only as a qualitative shape-closure diagnostic, not as an absolute validation of the total selection efficiency.}

As a separate internal consistency check, an independent final pass over the Delphes trees reproduces the frozen nominal event counts. For the massless sample, $89383$ events survive matching and $502$ pass the implemented signal-region selection, giving an implemented efficiency of $0.5616\%$.

Missing detector details cannot be promoted to calibrated experimental nuisance parameters. Instead, we define a narrower robustness test: one implemented requirement at a time is shifted around its nominal value while all others are held fixed. The variations are
\begin{align}
\ptg &: 45,\ 55~\gev,\\
\met &: 90,\ 110~\gev,\\
m_T &: 75,\ 85~\gev,
\end{align}
together with $\pm0.10$ changes of each angular threshold and the photon $|\eta|$ boundary. The treatment of an undefined jet-recoil angle is also toggled from the nominal fail convention to pass.

For a variation $v$, the quantity relevant to the mass extrapolation is the double ratio
\begin{equation}
D_v(m)=
\frac{A_v(m)/A_v(0)}{A_{\rm nom}(m)/A_{\rm nom}(0)}.
\label{eq:doubleratio}
\end{equation}
The minimum and maximum values of $D_v-1$ define the displayed selection-variation envelope. This construction tests migration near the public thresholds while preserving the massless normalization. It is not a one-standard-deviation detector systematic, and it is not added in quadrature to the Monte Carlo statistical uncertainty.

\section{Mass-dependent acceptance}
\label{sec:acceptance}

\begin{figure}[t]
\centering
\includegraphics[width=\columnwidth]{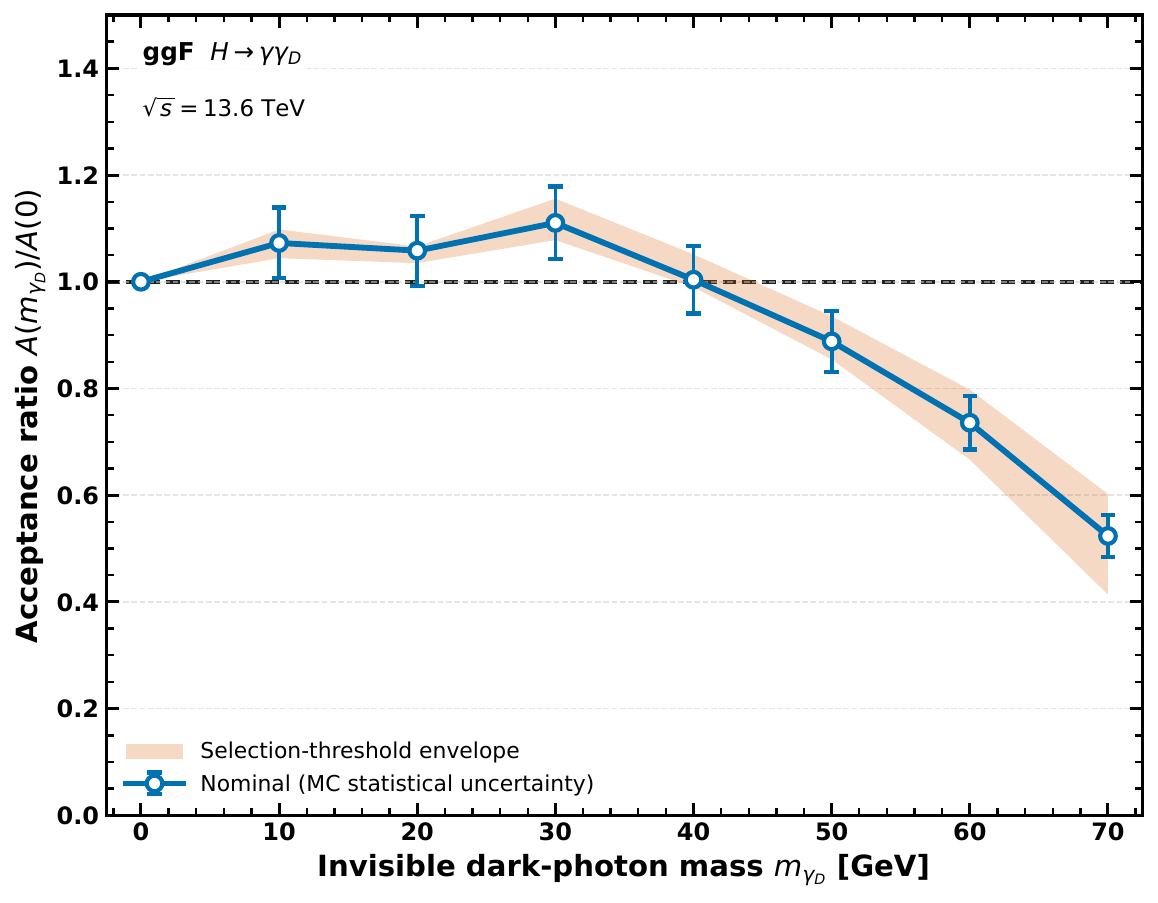}
\caption{\audit{Implemented ggF acceptance relative to the massless benchmark. Vertical bars show Monte Carlo statistical uncertainties. The orange band is the envelope of the one-at-a-time selection-threshold variations defined in Sec.~\ref{sec:validation}. It quantifies the stability of the \emph{relative} mass extrapolation and is not a calibrated ATLAS detector or theory systematic.}}
\label{fig:acceptance}
\end{figure}

\audit{Table~\ref{tab:acceptance} and Fig.~\ref{fig:acceptance} summarize the central simulation result. Within the available statistics, the acceptance ratio is consistent with a nearly mass-independent value through $m_{\gD}=40~\gev$. The modest upward fluctuations at $10$--$30~\gev$ are comparable to the $\sim6\%$ Monte Carlo uncertainty and should not be interpreted as a physical enhancement. The central value begins to decrease at $50~\gev$, with a substantially larger suppression at $60$ and $70~\gev$.}

\begin{table*}[t]
\caption{Matched and selected event counts, implemented efficiencies, acceptance ratios, Monte Carlo statistical uncertainties on the ratios, and selection-variation envelopes. The envelope at $m_{\gD}=0$ vanishes by construction because every variation is normalized to its own massless acceptance; it is not a statement of zero detector uncertainty on the anchor.}
\label{tab:acceptance}
\centering
\begin{ruledtabular}
\begin{tabular}{c r r c c c c c}
$m_{\gD}$ [GeV] & $N_{\rm match}$ & $N_{\rm sel}$ & $A$ [\%] & $R_A$ & $\delta_{\rm MC}$ [\%] & $\Delta_{\rm sel}^{-}$ [\%] & $\Delta_{\rm sel}^{+}$ [\%]\\
\hline
0  & 89383 & 502 & 0.5616 & 1.000 & 0.00 & 0.00 & 0.00\\
10 & 89626 & 540 & 0.6025 & 1.073 & 6.18 & -2.67 & +2.36\\
20 & 90007 & 535 & 0.5944 & 1.058 & 6.20 & -2.21 & +0.75\\
30 & 89475 & 558 & 0.6236 & 1.110 & 6.13 & -2.91 & +4.09\\
40 & 89399 & 504 & 0.5638 & 1.004 & 6.29 & -1.37 & +4.74\\
50 & 89395 & 446 & 0.4989 & 0.888 & 6.49 & -3.79 & +5.29\\
60 & 89765 & 371 & 0.4133 & 0.736 & 6.83 & -9.41 & +8.40\\
70 & 89102 & 262 & 0.2940 & 0.524 & 7.61 & -20.94 & +15.04\\
\end{tabular}
\end{ruledtabular}
\end{table*}

Numerically, the acceptance ratios at $50$, $60$, and $70~\gev$ are
\begin{equation}
R_A=0.888\pm0.058,
\quad
0.736\pm0.050,
\quad
0.524\pm0.040,
\label{eq:highmassRA}
\end{equation}
where the uncertainties are Monte Carlo statistical only. \rev{The decrease is in direct correspondence with the analytic scales in Table~\ref{tab:kinematics}: once $p_\gamma^*$ drops below the nominal photon threshold, accepted events must increasingly come from Higgs recoil, while the $m_T$ spectrum simultaneously moves toward the lower signal-region boundary.}

\begin{figure*}[t]
\centering
\includegraphics[width=0.94\textwidth]{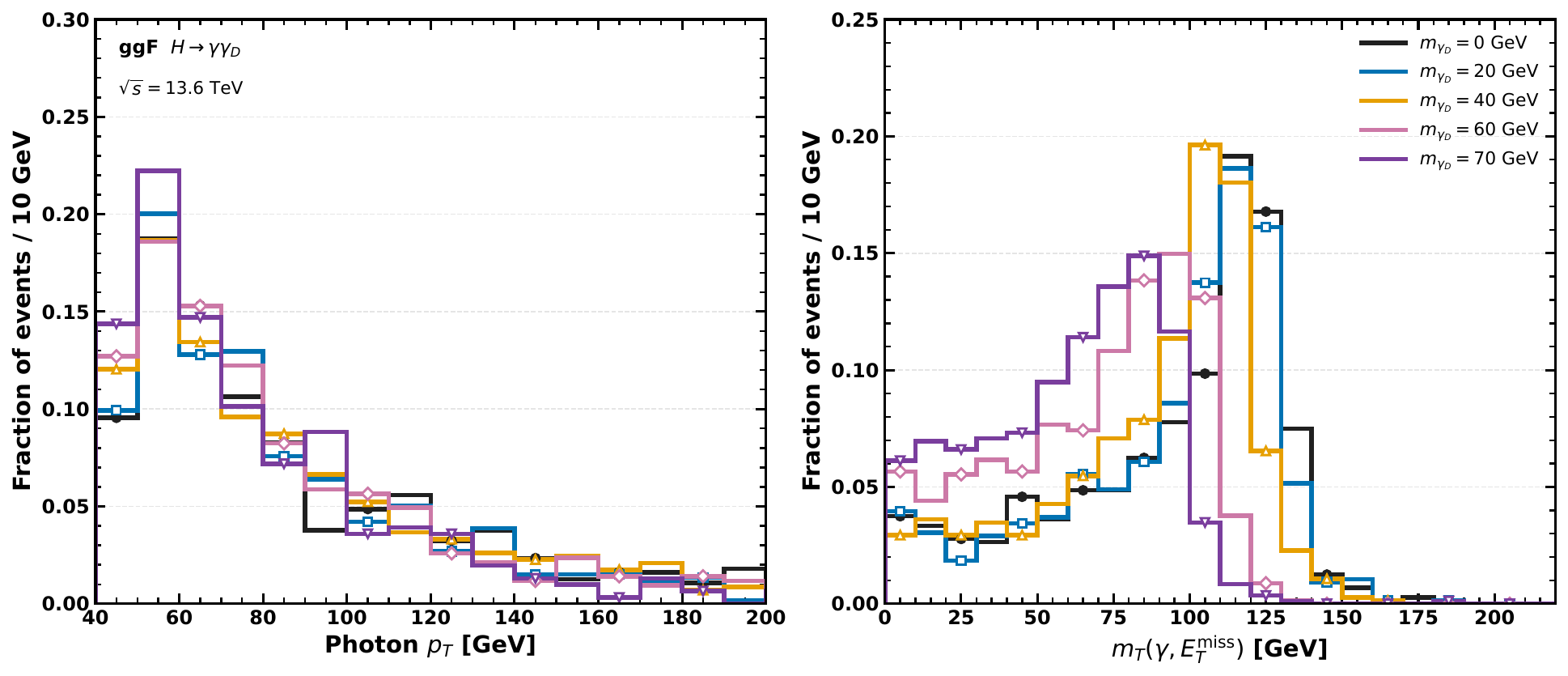}
\caption{\rev{Normalized $N\! -\!1$ kinematic distributions for representative dark-photon masses. Left: photon transverse momentum with the nominal $\ptg=50~\gev$ boundary indicated. Right: photon--missing-momentum transverse mass with the nominal $m_T=80~\gev$ boundary indicated. For each panel the corresponding threshold is omitted while the other common analysis requirements are retained. The progressive migration toward the thresholds explains why the upper-mass signal increasingly depends on Higgs recoil and why the relative acceptance becomes threshold sensitive.}}
\label{fig:nminusone}
\end{figure*}

\rev{The threshold origin of the acceptance loss is made more transparent by the $N\! -\!1$ distributions in Fig.~\ref{fig:nminusone}. In each panel the corresponding threshold is omitted while the remaining common requirements are retained, and each mass spectrum is normalized to unit area. The left panel shows the reconstructed photon-$p_T$ distribution and the right panel the transverse-mass distribution. As $m_{\gD}$ increases, the photon spectrum moves toward and below $50~\gev$, while the $m_T$ distribution accumulates near the $80~\gev$ boundary. The migration is already visible at $m_{\gD}=60~\gev$ and becomes pronounced at $70~\gev$, directly connecting the two-body scales in Table~\ref{tab:kinematics} to the reconstructed selection.}

\begin{figure*}[t]
\centering
\includegraphics[width=0.8\textwidth]{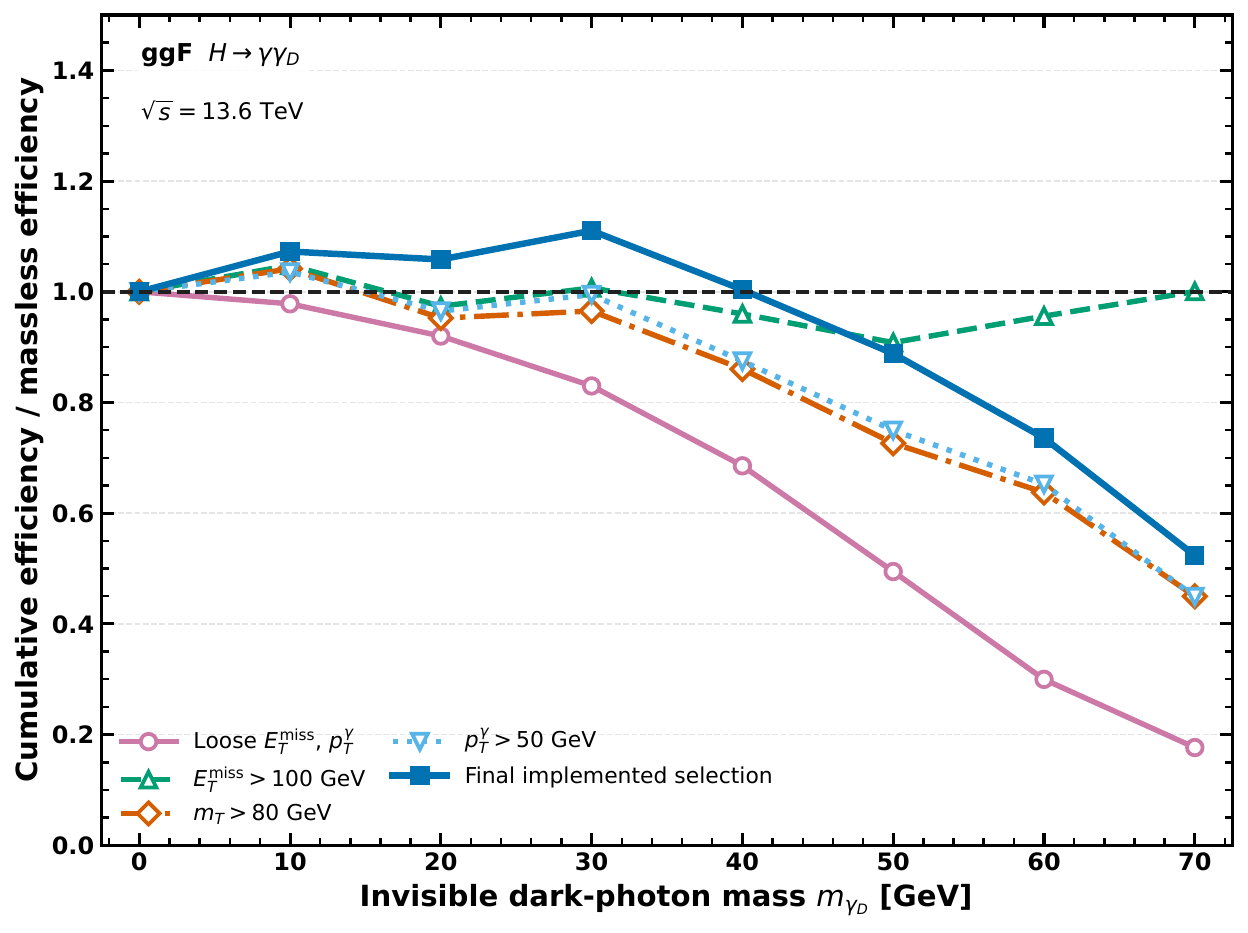}
\caption{\rev{Mass dependence of the cumulative efficiencies at selected stages of the implemented ggF analysis, each normalized to the corresponding efficiency at $m_{\gD}=0$. The comparison separates the early two-body kinematic suppression from the conditional effect of the harder $\met$, $m_T$, and photon-$p_T$ requirements. Because every curve is normalized to its own massless value, the plot is a diagnostic of mass dependence rather than an absolute detector-efficiency comparison.}}
\label{fig:cutimpact}
\end{figure*}

\rev{A complementary cumulative-efficiency diagnostic is shown in Fig.~\ref{fig:cutimpact}. Each curve is divided by its own massless efficiency, so the figure isolates how individual stages of the selection change with $m_{\gD}$. The loose photon-plus-missing-momentum stage already falls strongly with mass, reaching only about $0.18$ of its massless efficiency at $70~\gev$. The subsequent $\met>100~\gev$ requirement preferentially retains the recoil-enhanced high-mass subset and is therefore comparatively flat after normalization, whereas the $m_T$ and photon-$p_T$ requirements restore a strong high-mass suppression. The final implemented efficiency follows the acceptance ratio in Fig.~\ref{fig:acceptance}. This pattern shows that the endpoint behavior is a correlated kinematic migration through several thresholds, not an artifact of one isolated final cut.}

\audit{The largest absolute excursion of the selection-variation envelope is at most $5.3\%$ through $50~\gev$. At $60~\gev$ the envelope widens to approximately $-9.4\%/+8.4\%$, and at $70~\gev$ to $-20.9\%/+15.0\%$. The $m_T$ variation gives the largest excursion at the two highest masses. Under the specific variations tested here, the relative mass extrapolation is therefore comparatively stable through $50~\gev$ and becomes increasingly threshold dependent at $60$ and $70~\gev$. The $70~\gev$ result should be interpreted as an indicative public-recast estimate rather than a precision limit until the boosted-Higgs recoil is validated with higher-accuracy signal modeling.}

\section{Simplified likelihood and branching-fraction limits}
\label{sec:stats}

For a branching fraction $\BR\equiv\BR(H\to\gamma\gD)$, the expected ggF signal count is taken to be
\begin{equation}
 s(m_{\gD},\BR)
 =166\,R_A(m_{\gD})\frac{\BR}{0.01}.
\label{eq:signalcount}
\end{equation}
\audit{The relative signal-normalization uncertainty combines the public uncertainty on the massless ggF anchor, $24/166$, with the Monte Carlo uncertainty on $R_A$. Treating these two contributions as independent, we use
\begin{equation}
 \delta_s^2(m_{\gD})=
 \left(\frac{24}{166}\right)^2+
 \left(\frac{\sigma_{R_A}(m_{\gD})}{R_A(m_{\gD})}\right)^2,
 \label{eq:signalunc}
\end{equation}
consistent with standard first-order uncertainty propagation \cite{PDG2024}. The selection-variation envelope is evaluated separately as discrete low- and high-acceptance scenarios rather than as a Gaussian nuisance.}

\audit{Following standard profile-likelihood constructions used in collider searches \cite{Cowan2011,CranmerHistFactory2012}, we model the inclusive signal region with a single Poisson count and Gaussian-constrained effective background and signal-normalization nuisance parameters,}
\begin{align}
\mathcal{L}(\BR,\theta_b,\theta_s)
&=\mathrm{Pois}\!\left[n\,\middle|\,
 b+\sigma_b\theta_b+s(m_{\gD},\BR)(1+\delta_s\theta_s)\right]
\nonumber\\
&\quad\times G(\theta_b;0,1)G(\theta_s;0,1),
\label{eq:likelihood}
\end{align}
with
\begin{equation}
 n=11327,
 \qquad b=11298,
 \qquad \sigma_b=101.
\end{equation}
\audit{Here $G(\theta;0,1)$ denotes a unit-normal constraint and $\delta_s$ is defined in Eq.~\eqref{eq:signalunc}. The branching fraction is constrained to $\BR\geq0$. Expected limits are evaluated with the background-only Asimov data set \cite{Cowan2011}. We use a one-sided profile-likelihood test statistic and the asymptotic $\CLs$ prescription to derive 95\% confidence-level upper limits \cite{Read2002,Cowan2011}.}

This likelihood is deliberately simpler than the experimental one. The published post-fit background in Eq.~\eqref{eq:publiccounts} originates from a simultaneous multiregion fit that includes signal-region and control-region shape information, correlated systematic uncertainties, and floating background normalizations. Equation~\eqref{eq:likelihood} compresses that information into an effective one-bin Gaussian background constraint and therefore cannot reproduce the full ATLAS likelihood.

\audit{A useful closure check is obtained by replacing the ggF signal anchor in Eq.~\eqref{eq:signalcount} by the sum of the published massless ggF, VBF, $WH$, and $ZH$ signal yields while keeping the same one-bin statistical construction. This gives an expected limit of $1.195\%$ and an observed limit of $1.280\%$, compared with the official Run~3 values of $1.2\%$ and $1.4\%$, respectively \cite{ATLASRun3DarkPhoton2026}. The expected limit differs by about $0.4\%$ relative to the official value, whereas the observed limit differs by about $8.6\%$. The agreement in the expected limit provides a useful normalization cross-check, while the larger observed difference illustrates the information lost when the full control-region and shape likelihood is compressed to one effective count.}

\begin{figure}[t]
\centering
\includegraphics[width=\columnwidth]{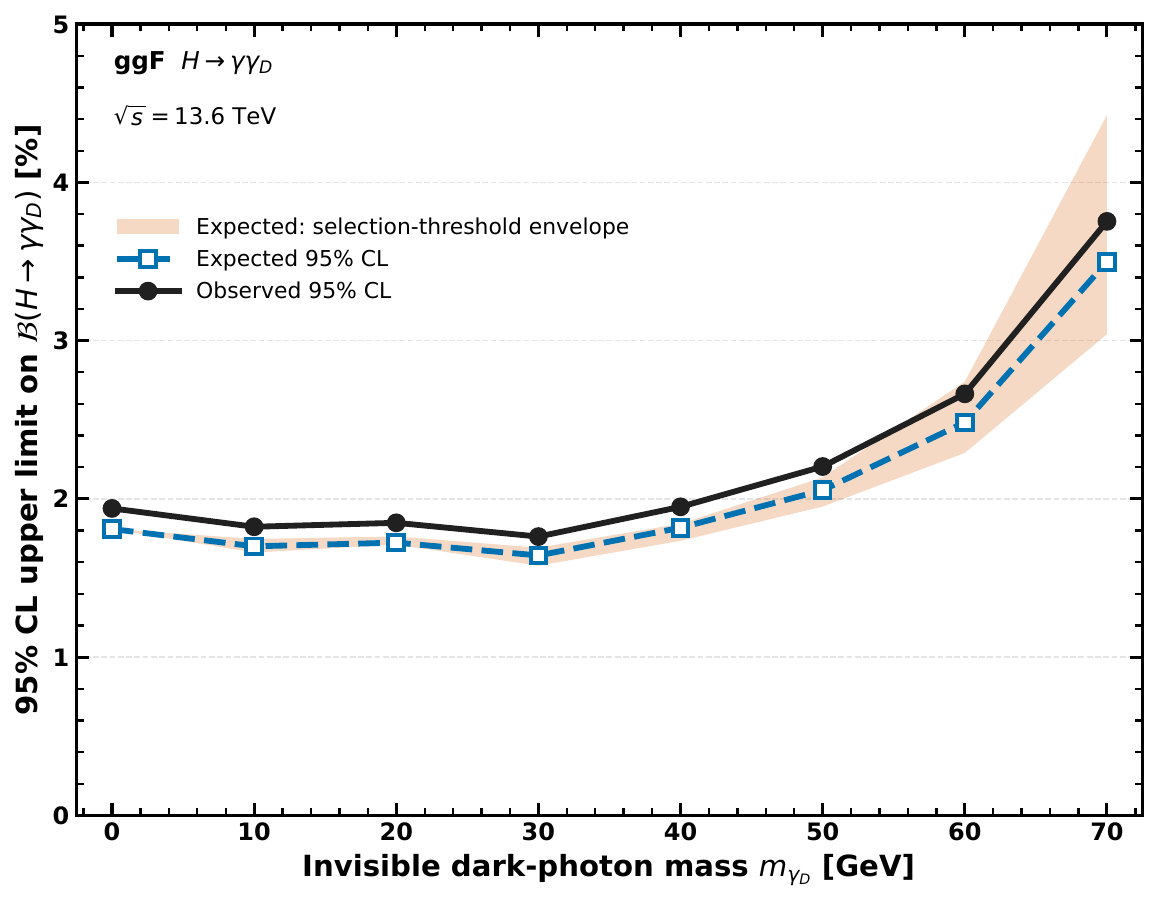}
\caption{\audit{Expected and observed simplified 95\% $\CLs$ upper limits on $\BR(H\to\gamma\gD)$ as functions of $m_{\gD}$. The orange band is the expected-limit bracket obtained by propagating the selection-threshold envelope. It is not a calibrated experimental uncertainty. The result is a ggF-only one-bin public reinterpretation and should not be quoted interchangeably with the official inclusive ATLAS limit.}}
\label{fig:limits}
\end{figure}

Table~\ref{tab:limits} and Fig.~\ref{fig:limits} give the resulting ggF-only mass-dependent limits. The central expected limit is approximately flat below $40~\gev$ within the resolution of the acceptance scan. The numerically smallest value, $1.642\%$ at $30~\gev$, coincides with a positive Monte Carlo fluctuation of the acceptance and should not be interpreted as a physically preferred mass. The limit weakens to $2.055\%$ at $50~\gev$, $2.483\%$ at $60~\gev$, and $3.498\%$ at $70~\gev$. The observed values follow the same pattern because the public inclusive signal-region count lies close to the fitted background prediction.

\begin{table*}[t]
\caption{Simplified 95\% $\CLs$ upper limits on $\BR(H\to\gamma\gD)$ for ggF production. Bracketed intervals propagate the discrete selection-variation envelope only; they do not include an independently calibrated systematic for ATLAS-only requirements or higher-order modeling of the ggF recoil.}
\label{tab:limits}
\centering
\begin{ruledtabular}
\begin{tabular}{c c c c c}
$m_{\gD}$ [GeV] & Expected [\%] & Expected selection bracket [\%] & Observed [\%] & Observed selection bracket [\%]\\
\hline
0  & 1.809 & [1.809, 1.809] & 1.940 & [1.940, 1.940]\\
10 & 1.700 & [1.661, 1.747] & 1.824 & [1.782, 1.874]\\
20 & 1.723 & [1.711, 1.762] & 1.849 & [1.835, 1.890]\\
30 & 1.642 & [1.578, 1.692] & 1.762 & [1.692, 1.814]\\
40 & 1.817 & [1.735, 1.843] & 1.950 & [1.861, 1.977]\\
50 & 2.055 & [1.952, 2.136] & 2.204 & [2.093, 2.291]\\
60 & 2.483 & [2.290, 2.741] & 2.664 & [2.457, 2.940]\\
70 & 3.498 & [3.041, 4.425] & 3.753 & [3.263, 4.747]\\
\end{tabular}
\end{ruledtabular}
\end{table*}

\section{Interpretation and limitations}
\label{sec:discussion}

\audit{The central phenomenological result is that the massless benchmark provides a good approximation to the implemented ggF acceptance for invisible masses up to about $40~\gev$, but not throughout the full kinematically allowed two-body range. In the simulation, the loss of acceptance above $50~\gev$ is traced to the interplay between the changing two-body decay kinematics and the fixed event-selection thresholds. It therefore cannot be captured by applying only the partial-width phase-space factor in Eq.~\eqref{eq:phasewidth}.}

The two relevant mechanisms are visible already at the Born level. First, $p_\gamma^*$ decreases with $m_{\gD}$ and crosses the nominal reconstructed photon threshold near $56~\gev$. Second, the conventional massless-proxy $m_T$ edge approaches the $80~\gev$ threshold. Both effects make the high-mass selection dependent on transverse recoil of the Higgs boson. \rev{The $N\! -\!1$ shapes in Fig.~\ref{fig:nminusone} demonstrate the corresponding reconstructed spectral migration directly, while the cumulative-efficiency ratios in Fig.~\ref{fig:cutimpact} show how that migration propagates through the implemented selection. Together with the representative cutflow in Appendix~\ref{app:cutflow}, these diagnostics make the origin of the high-mass acceptance loss substantially more transparent than a final-efficiency curve alone.}

\audit{The ratio construction in Eq.~\eqref{eq:ratio} reduces the dependence of the result on the absolute fast-simulation efficiency. Photon reconstruction, jet response, and several details of the approximate object selection are common to all masses and can therefore partially cancel against the massless benchmark. The size of the selection-variation envelope through $50~\gev$ is consistent with such partial cancellation under the variations tested here.}

This cancellation is not exact. The missing ATLAS-only requirements can acquire indirect mass dependence through correlations with $\ptg$, $\met$, jet recoil, or event topology. The present study cannot calibrate such effects. For this reason, the ratio should be interpreted as the recastable mass dependence under a clearly stated factorization assumption, not as a full detector-level transfer function.

\audit{A key unquantified theory effect for the upper mass points is the modeling of the boosted ggF recoil. The present hard process is leading order with matched zero- and one-parton matrix elements, while the experimental signal modeling uses higher perturbative accuracy and a more refined Higgs transverse-momentum description \cite{ATLASRun3DarkPhoton2026,Hamilton2013,deFlorian2017,Karlberg2024}. At low $m_{\gD}$ this difference is less consequential because the two-body decay kinematics more readily satisfy the photon threshold. At $60$--$70~\gev$, by contrast, the selected sample increasingly consists of events in which QCD recoil boosts the Higgs system.}

\audit{The cut-impact diagnostic sharpens this conclusion: because the accepted high-mass sample is selected from a recoil-enhanced subset, a key missing theory validation is the modeling of the Higgs transverse-recoil spectrum rather than another variation of the final analysis threshold.} A precision extension should therefore compare $R_A(m)$ under variations of renormalization and factorization scales, merging parameters, parton-shower recoil schemes, and ideally NLO+PS or NNLOPS Higgs-production samples. Importantly, the higher-order study should be performed as a mass-dependent ratio rather than as a generic inclusive ggF uncertainty: the issue is whether the recoil modeling changes different $m_{\gD}$ points by different amounts after the low-threshold selection.

The Run~3 signal region contains a substantial non-ggF component in the massless benchmark. VBF and associated production have different recoil and jet topologies and can therefore exhibit a different mass dependence from ggF. In particular, additional hard objects in VBF, $WH$, and $ZH$ can alter the boost of the Higgs boson and the behavior of the angular and jet-based selections. The published Run~2 studies already demonstrate that these production modes have distinct experimental acceptances \cite{CMSZH2019,CMSVBF2021,ATLASVBF2022,ATLASZH2023,ATLASCombination2024}.

For this reason, an all-production-mode massive-dark-photon limit cannot be obtained by simply multiplying the ggF result by the massless inclusive signal fraction. The correct extension requires explicit massive samples for each production mode, followed by a combined statistical model. The present ggF result remains useful because ggF supplies the largest signal component and cleanly exposes the threshold physics of the dominant channel.

\audit{A further distinction from the massless benchmark concerns the microscopic interpretation. For an exactly massless dark photon, the gauge kinetic terms can be diagonalized by field redefinitions, while observable effects depend on the matter charge assignments and on interactions connecting the visible and dark sectors. For a massive vector, kinetic mixing generally leads to physical couplings to the Standard Model electromagnetic current \cite{Holdom1986,Dienes1997,Fabbrichesi2020,BentoHaberSilva2024}. Consequently, a massive dark photon can decay visibly unless the mixing is sufficiently small or invisible dark-sector decay modes dominate. The present branching-fraction limits are therefore most model independent when stated for a detector-invisible spin-one state rather than directly as bounds on a universal kinetic-mixing parameter.}

A complete ultraviolet reinterpretation would require at least the effective $H\gamma\gD$ coupling, the visible kinetic mixing, the dark gauge coupling, the masses of any lighter dark states, and the resulting detector-frame lifetime. Existing invisible dark-photon searches in fixed-target and flavor-factory experiments constrain some of these parameter combinations at lower masses \cite{BaBarInvisible2017,NA64Invisible2018,NA64MissingEnergy2019,NA64Latest2023}. Visible-resonance searches probe complementary lifetime and branching regimes \cite{BaBarVisible2014,LHCbDarkPhoton2020}. Those bounds are complementary rather than directly equivalent to the Higgs branching-fraction limits reported here.

The numerical closure at $m_{\gD}=0$ is reassuring but should not be overinterpreted. The official Run~3 result derives sensitivity from a simultaneous binned likelihood over the signal region and multiple control regions, with correlated experimental, theoretical, and data-driven nuisance parameters \cite{ATLASRun3DarkPhoton2026}. In contrast, Eq.~\eqref{eq:likelihood} treats the published post-fit background as if it were an external Gaussian-constrained quantity. The resulting observed limit can differ more visibly than the expected limit because the treatment of the actual data fluctuations and control-region constraints is not reproduced.

A stronger public recast would require a released simplified likelihood or full statistical model. If such information becomes available, the same mass-dependent signal templates could be propagated through the official binning. The current selected Monte Carlo populations at high mass are sparse in the upper $m_T$ bins, however, so higher-statistics samples would also be required before a four-bin signal-shape reinterpretation becomes numerically stable.

\audit{Within these limitations, the present study supports three statements. First, the mass dependence is weak below approximately $40~\gev$ for the ggF signal under the implemented selection. Second, the degradation above $50~\gev$ is traced to kinematic migration relative to the fixed $\ptg$ and $m_T$ thresholds. Third, the available public information permits this acceptance ratio to be translated into an approximate branching-fraction sensitivity, provided the result is explicitly identified as ggF only and based on a simplified statistical model.}

\audit{The analysis therefore complements, rather than replaces, the experimental search. Its main utility is to provide a reproducible public mapping of the ggF acceptance away from the published massless Run~3 benchmark and to identify the mass range in which additional signal-model accuracy becomes important.}

\section{Conclusions}
\label{sec:conclusions}

\audit{We have presented a public reinterpretation of the ATLAS Run~3 $H\to\gamma+\met$ search for a massive detector-invisible dark photon in Higgs production through ggF.} Eight masses between $0$ and $70~\gev$ were processed through a common hard-event, matched parton-shower, fast-detector, and analysis chain. The absolute signal normalization was inherited from the published massless ggF yield, while the simulation was used only to determine the relative acceptance $R_A(m_{\gD})$.

\audit{The central result is that the relative acceptance $R_A(m_{\gD})$ is statistically compatible with a flat mass dependence through $m_{\gD}=40~\gev$. It then falls to $0.888$, $0.736$, and $0.524$ at $50$, $60$, and $70~\gev$, respectively.} The behavior is explained by two-body Higgs-decay kinematics: the photon rest-frame momentum approaches and then falls below the $50~\gev$ analysis threshold, while the conventional massless-invisible transverse-mass distribution is compressed toward $m_T=80~\gev$. \rev{The $N\! -\!1$ spectra and cumulative cut-impact ratios verify this mechanism at reconstructed level and show that the suppression is distributed across correlated kinematic requirements rather than being caused by one anomalous cut.} The upper mass points therefore rely increasingly on recoil of the Higgs system against QCD radiation.

\audit{Using the public signal-region count and post-fit background, the one-bin asymptotic $\CLs$ construction gives expected (observed) ggF-only limits of $1.64\%$ ($1.76\%$) at $m_{\gD}=30~\gev$ and $3.50\%$ ($3.75\%$) at $70~\gev$. The local minimum at $30~\gev$ follows the positive Monte Carlo fluctuation of the acceptance and is not interpreted as a preferred mass. Within the Monte Carlo precision, the low-mass sensitivity is approximately flat through $40~\gev$. The one-at-a-time selection-variation envelope has a largest absolute excursion below $5.3\%$ through $50~\gev$, increasing to about $9.4\%$ at $60~\gev$ and $20.9\%$ at $70~\gev$.}

\audit{The reliability of the mass extrapolation is therefore not uniform across the scan. Under the implemented threshold variations, the $0$--$50~\gev$ region is the least sensitive to the tested selection changes. Sensitivity to those choices increases at $60~\gev$ and becomes substantial at $70~\gev$. We therefore treat the $70~\gev$ point as indicative rather than a precision result until the boosted ggF recoil is tested with higher-order signal samples and increased Monte Carlo statistics. Explicit massive samples for VBF, $WH$, and $ZH$ production are additionally required before an all-production-mode massive-dark-photon limit can be constructed.}

\audit{More broadly, the analysis illustrates why the massless $H\to\gamma+\text{invisible}$ benchmark cannot automatically be extended to a massive invisible daughter by a decay phase-space rescaling. Once fixed experimental thresholds become comparable to the daughter kinematics, detector acceptance becomes an essential part of the interpretation. The mass-dependent results reported here provide a reproducible public reference for future phenomenological studies and for ultraviolet models predicting a massive detector-invisible spin-one state in Higgs decay.}

\section*{Acknowledgments}

We acknowledge the use of AI-assisted tools, in particular ChatGPT, to improve the English language, grammar, and readability of the manuscript. The authors remain fully responsible for the scientific content, analysis, and conclusions presented in this work.

\appendix

\section{Representative cutflows}
\label{app:cutflow}

\begin{table*}[t]
\caption{Representative cumulative cutflows. Requirements in parentheses are not applied in the fast simulation and do not remove events.}
\label{tab:cutflow}
\centering
\begin{ruledtabular}
\begin{tabular}{l r r r}
Cumulative requirement & $m_{\gD}=0$ & $m_{\gD}=20~\gev$ & $m_{\gD}=70~\gev$\\
\hline
Matched events & 89383 & 90007 & 89102\\
$\met>50~\gev$ and $\ptg>45~\gev$ & 27505 & 25484 & 4846\\
(Analysis trigger) & 27505 & 25484 & 4846\\
$\met>100~\gev$ & 1505 & 1476 & 1501\\
$m_T>80~\gev$ & 1126 & 1080 & 505\\
Exactly one baseline photon & 1122 & 1078 & 501\\
$\ptg>50~\gev$ & 960 & 933 & 430\\
Lepton veto & 937 & 910 & 419\\
At most three central jets & 803 & 792 & 359\\
$b$-jet veto & 760 & 758 & 347\\
(Vertex BDT, $\met$ significance, no-JVT $\met$) & 760 & 758 & 347\\
$\Delta\phi(\met,-\gamma)>1.25$ & 662 & 683 & 338\\
$\Delta\phi(\met,-\sum\bm p_T^{\rm jets})<0.75$ & 644 & 659 & 335\\
$\Delta\phi(j_1,j_2)<2.5$ & 586 & 605 & 301\\
$|\eta_\gamma|<1.75$ & 502 & 535 & 262\\
\end{tabular}
\end{ruledtabular}
\end{table*}

Table~\ref{tab:cutflow} reports the cumulative implemented cutflow at three representative masses. Parentheses denote public ATLAS requirements that are not applied in the fast simulation and whose counts are therefore carried forward unchanged. The comparison localizes the high-mass acceptance loss. At $70~\gev$, the loose $\met>50~\gev$ and $\ptg>45~\gev$ stage retains only $4846$ of $89102$ matched events, compared with $27505$ of $89383$ for the massless sample. The later $\met$, $m_T$, and photon requirements reduce the $70~\gev$ sample to $430$ events before the remaining veto and angular selections.

\section{Selected transverse-mass populations}
\label{app:mtbins}

For completeness, Table~\ref{tab:mtcounts} gives the selected unweighted Monte Carlo populations in the four transverse-mass intervals used by the public Run~3 analysis. The upper bins become sparsely populated at high $m_{\gD}$, which is why the present public likelihood is kept inclusive rather than constructed from these simulated bins.

\begin{table}[t]
\caption{Selected simulated events in the four $m_T$ intervals. These are unweighted Monte Carlo populations after the implemented selection, not predicted ATLAS yields.}
\label{tab:mtcounts}
\centering
\begin{ruledtabular}
\begin{tabular}{c r r r r}
$m_{\gD}$ [GeV] & 80--110 & 110--140 & 140--200 & $\geq200$\\
\hline
0  & 172 & 313 & 17 & 0\\
10 & 210 & 304 & 25 & 1\\
20 & 215 & 302 & 17 & 1\\
30 & 296 & 248 & 14 & 0\\
40 & 291 & 201 & 11 & 1\\
50 & 322 & 118 & 5 & 1\\
60 & 333 & 38  & 0 & 0\\
70 & 250 & 11  & 1 & 0\\
\end{tabular}
\end{ruledtabular}
\end{table}

\bibliographystyle{apsrev4-2}
\bibliography{references}

\end{document}